\documentclass{epl}

\usepackage{epsfig}

\title{Mean first passage time analysis reveals rate-limiting steps,
parallel pathways and dead ends in a simple model of protein
folding}

\shorttitle{Mean first passage time analysis of protein folding}

\author{M. A. Micheelsen\inst{1}, C. Rischel\inst{1,2}\footnote{Corresponding author, E-mail rischel@nbi.dk},
 J. Ferkinghoff-Borg\inst{1,3}, R. Guerois\inst{3} and L. Serrano\inst{3}}

\shortauthor{M. A. Micheelsen \etal}

\institute{
  \inst{1} Niels Bohr Institute, University of Copenhagen, Blegdamsvej 17, DK-2100 K\o benhavn \O, Denmark\\
  \inst{2} Department of Mathematics and Physics, Royal Veterinary and Agricultural University
  Thorvaldsensvej 40, DK-1871 Frederiksberg C,Denmark\\
  \inst{3} EMBL Heidelberg, Meyerhofstrasse 1, D-69117 Heidelberg, Germany
}

\pacs{87.15.Aa}{Theory and modeling; computer simulation}
\pacs{82.20.Fd}{Stochastic and trajectory models; other theories
and models}

\begin{document}

\maketitle

\begin{abstract}
We have analyzed dynamics on the complex free energy landscape of
protein folding in the FOLD-X model, by calculating for each state
of the system the mean first passage time to the folded state. The
resulting kinetic map of the folding process shows that it
proceeds in jumps between well-defined, local free energy minima.
Closer analysis of the different local minima allows us to reveal
secondary, parallel pathways as well as dead ends.
\end{abstract}

Proteins are polymers with several hundred degrees of freedom, so
it is natural to think of protein folding as (diffusive) motion on
a complex potential energy surface. This {\em landscape} picture
has become popular in the wider protein folding community during
the last decade \cite{Wolynes,Onuchic}, and clearly invites the
question of the nature of the surface: ruggedness, presence of
kinetic traps, funnels etc. One problem with answering such
questions is that the concept of folding rate, used to describe
simple two-state kinetics, can not be generalized to a property
that is defined for each configuration of the protein, so it is
not useful for describing local kinetics. In this Letter, we use
the {\em inverse} rate, the mean first passage time (MFPT) which
{\em can} be calculated for each configuration \cite{MFPT1}, to
analyze the kinetics of folding. Our results clearly reveal
kinetic barriers that can not be predicted from the
one-dimensional projection of the free energy landscape, and allow
us to construct an improved reaction coordinate.

Workable, detailed models of protein folding energetics and
kinetics do not exist. Instead, we work with the FOLD-X model
\cite{FoldX1}, which is rather simplified, but takes experimental
knowledge into account. FOLD-X and similar models
\cite{Galzitskaya, AlmBaker, MunozEaton} are based on the
observation that the folding rate of small proteins is correlated
with the entropy cost of ordering the chain in a near-native
geometry, implying that this happens before the transition state.
Non-native interactions are hereby rendered less likely, and the
models only consider interactions present in the folded structure.
In these approximations, any residue can adopt only two states;
folded/ordered or unfolded/disordered. The conformation a folded
residue is always set to its native state, whereas a contiguous
sequence of unfolded residues is treated as a disordered loop.

In FOLD-X, the total free energy, $G$, of a specific protein state
(defined by the binary sequence of folded/unfolded residues) is
the sum of the interactions between the atoms of the folded
residues plus terms accounting for the loss of chain entropy:
\begin{equation}
 G=
W_{vdw}G_{vdw}+W_{solvH}G_{solvH}+W_{solvP}G_{solvP}+G_{hbond}+G_{el}
+T\left(W_{mc}S_{mc}+W_{sc}S_{sc}+S_{loop}\right) \label{FOLD-X-G}
\end{equation}
Here, $G_{vdw}$ is the sum of the van der Waals contributions of
all atoms of the folded residues. $G_{solvH}$ and $G_{solvP}$ is
the difference in solvation energy upon folding of apolar and
polar groups, respectively. $G_{hbond}$ is the free energy
difference between the formation of intra-molecular hydrogen-bonds
compared to the inter-molecular formation of hydrogen-bonds with
solvent. $G_{el}$ is the electrostatic contribution of charged
groups interactions. $S_{mc}$ and $S_{sc}$ are the sum of the
entropy-loss associated with the ordering of respectively the
backbone and the side-chain of each folded residue. Finally,
$S_{loop}$ is the sum of the entropy cost associated with the
closure of the disordered loops connecting the stretches of the
folded residues. The strength of the various interactions
($G_{vdw}$, $G_{solvH}$, $G_{solvP}$, $G_{el}$) and the entropy
cost for ordering a residue ($S_{mc}$, $S{sc}$) are scaled the
atomic occupancies \cite{solvent} to take the effect of solvent
expose into account \cite{FoldX1}. The terms $W_{vdw}$,
$W_{solvH}$, $W_{solvP}$, $W_{mc}$ and $W_{sc}$ in eq.
(\ref{FOLD-X-G}) are weighting factors applied to the raw
energy/entropy terms. These weights have been obtained from a
calibration against a comprehensive database of protein mutants
\cite{FoldX2}. The details of the different energy and entropy
terms can be found in \cite{FoldX1,FoldX2}.

In order to reduce the size of the state space, only states with
one or two segments of ordered residues are considered. A
comparison with an unrestricted Monte Carlo based sampling shows,
that the two-segment approximation captures the thermodynamic
behavior of small single domain proteins quite well \cite{FoldX3}.
We use the local kinetics of the energy model: in each step, the
protein can add or remove a residue to an end of one of the
ordered segments (including the possibility to remove a
one-residue segment) or create a new one-residue segment if zero
or one ordered segments are present. The move probabilities $P$
between states $i$ and $j$ are chosen so they fulfill the detailed
balance condition
\begin{equation}
\frac{P(i\rightarrow j)}{P(j\rightarrow i)} =
\exp(-(G_j-G_i)/k_BT) \label{eq:MicroRev}
\end{equation}
This is achieved by using
\begin{equation}
P(i\rightarrow j) = {\rm max}(M_i,M_j)^{-1} \, {\rm
max}(1,\exp(-(G_j-G_i)/k_BT)) \label{eq:ProbDef}
\end{equation}
where $M_i$ and $M_j$ are the number of moves available from state
$i$ and $j$, respectively. This expression clearly fulfills
(\ref{eq:MicroRev}), while also ensuring that the $M_i$ move
probabilities out of state $i$ add up to at most 1. The
probability for remaining is $P(i\rightarrow i) = 1-\sum_{j\neq i}
P(i\rightarrow j)$.

Since the kinetics are ergodic all trajectories starting in a
particular state $i$ will sooner or later end up in any other
given state $n$. We can define the average time (number of steps)
before this happens, the mean first passage time $\tau_i(n)$. If
we advance one step along these trajectories we clearly get one
step closer to the passage through state $n$. Since we advance by
the move probabilities $P$ the MFPT obeys \cite{MFPT1}
\begin{equation}
\tau_i(n) = \sum_j P(i\rightarrow j) \tau_j(n) + \Delta t
\label{eq:MFPT1}
\end{equation}
where $\Delta t$ is the length of the time step. For $i=n$ the
special condition $\tau_n(n)=0$ must hold. If we define the matrix
$T(n)$ as follows:
\begin{equation}
T_{ij}(n) = \left\{
  \begin{array}{ll}
    P(i\rightarrow j) & i\neq n \\
    0 & i=n
  \end{array}
  \right.
\label{eq:Wdef}
\end{equation}
and the vector $\beta(n)$ by
\begin{equation}
\beta_i (n) = \left\{
  \begin{array}{ll}
    \Delta t & i \neq n \\
    0 & i=n
  \end{array} \right.
\label{eq:betadef}
\end{equation}
the condition (\ref{eq:MFPT1}) and the special case for $i=n$ can
be summarized in a matrix equation for the MFPT vector $\tau(n)$:
\begin{equation}
[T(n)-1] \tau (n) = - \beta
  \label{eq:MFPT2}
\end{equation}
We solve this equation numerically using the  {\tt linbcg} sparse
matrix routine \cite{NumRec} and with $n$ set to the completely
folded state.

\begin{figure}
\centering \epsfig{file=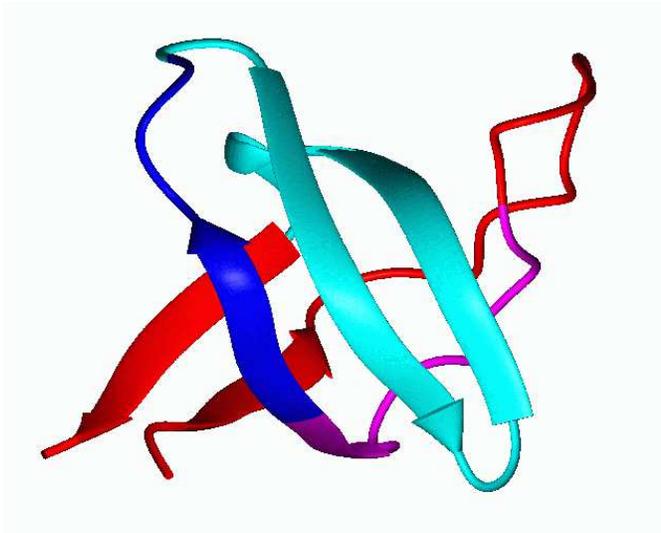,width=250pt}
 \caption{Structure of a
Src Homology 3 domain, PDB-file 1shg. The colour coding shows
different stages in folding in the FOLD-X model, as described in
the main text. The figure was rendered by MolMol \cite{MolMol}.}
\label{f.1}
\end{figure}

\begin{figure}
\epsfig{file=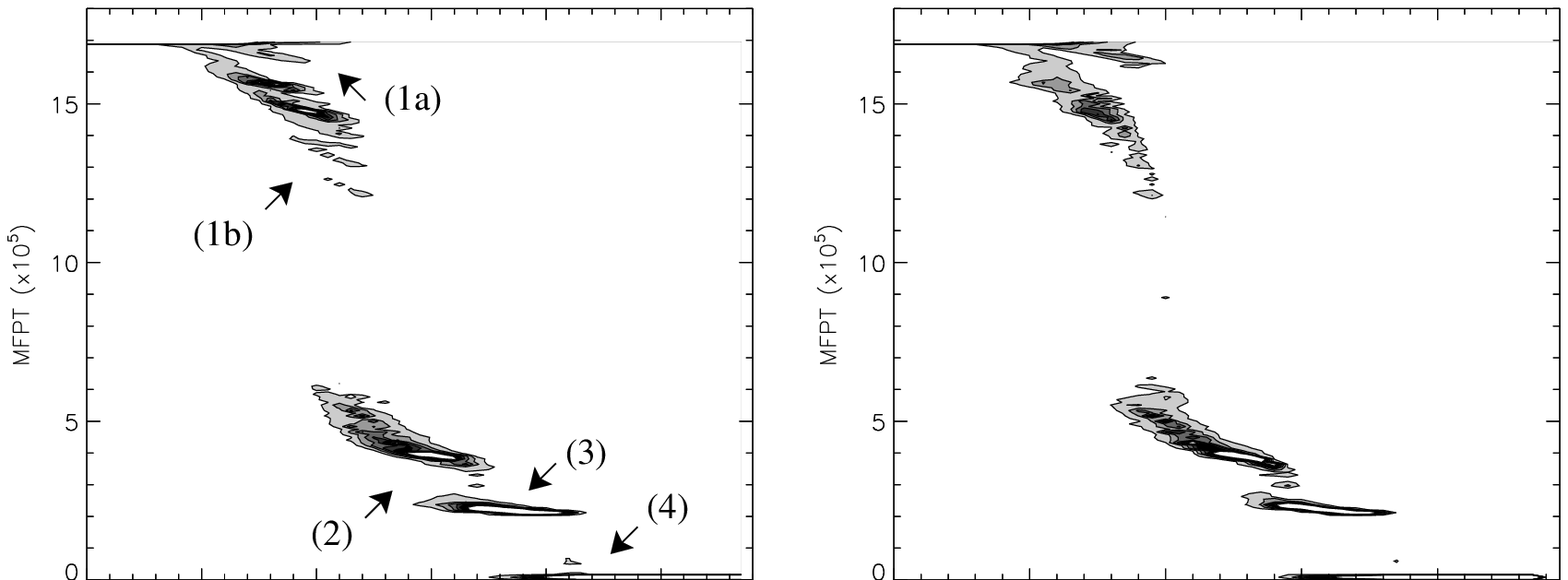,width=\textwidth}
\epsfig{file=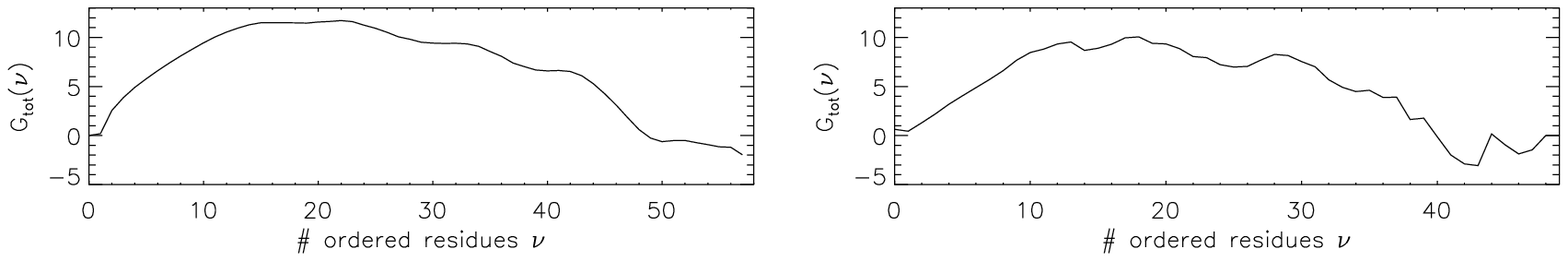,width=\textwidth} \caption{Left, Upper
panel: Contour plot of the Boltzmann weighted probability of
different mean first passage times (ordinate) as function of the
number of ordered residues (abscissa) in the SH3 domain (pdb file
1shg), with dark colour indicating high probability. Lower panel:
Total free energy as function of number of folded residues. Right:
Same as left but not counting the 3 N-terminal residues and 6
C-terminal residues.} \label{f.2}
\end{figure}

We here present results from the calculation on an SH3 domain,
which has 57 amino acid residues and a total of 425924 different
states in the 2-segment approximation. The x-ray structure is
shown in Figure \ref{f.1} (Protein Data Bank file 1shg
\cite{SH3struct}). In order to present the results in condensed
form we sum the Boltzmann probability over states with same MFPT
$\tau$ and number of ordered residues $\nu$
\begin{equation}
\Gamma (\nu,\tau) = \sum_j \exp (-G_j/k_B T) \delta(\tau-\tau_j)
\delta(\nu - \nu_j) \label{eq:GammaDef}
\end{equation}
where $\delta$ is the Kronecker delta. We also calculate the
one-dimensional free energy
\begin{equation}
G_{\rm tot} (\nu) = - k_B T \log [\sum_\tau \Gamma(\nu,\tau)
]\label{eq:GtotDef}
\end{equation}
The upper panels in Figure \ref{f.2} show contour plots of
$\Gamma(\nu,\tau)/ \exp(-G_{\rm tot} (\nu)/k_B T)$. The division
with the relative probability for a given $\nu$ is done in order
also to show the kinetics at the top of the barrier. The lower
panels show total free energy as function of the number of ordered
residues, with the free energy in the unfolded state normalized to
zero.

In the left part of Figure \ref{f.2} the reaction coordinate $\nu$
takes ordering of all residues into account, as is usual with this
type of model \cite{FoldX1,Galzitskaya,AlmBaker,MunozEaton}. The
free energy is seen to be a smooth function of the number of
ordered residues, with minima at the unfolded and folded states
and a maximum between them (the transition state). As expected,
the MFPT to the folded state is high for states with few ordered
residues and low for states with almost all residues ordered. Less
expected is the formation of 'islands' with gaps between them. The
gaps show the presence of particular critical events that in one
step reduce the MFPT significantly. This suggests that the smooth
variation of the free energy seen in the lower panel might be a
poor picture of the actual energetics.

Surprisingly, the relation between MFPT and residue ordering is
not monotonous; for many states it is possible to find another
state with fewer ordered residues but lower MFPT. We show in the
following that this behaviour can have three reasons:
\begin{enumerate}
\item Parallel folding pathways \item Poor choice of reaction
coordinate \item Dead ends
\end{enumerate}

\begin{figure}
\centering \epsfig{file=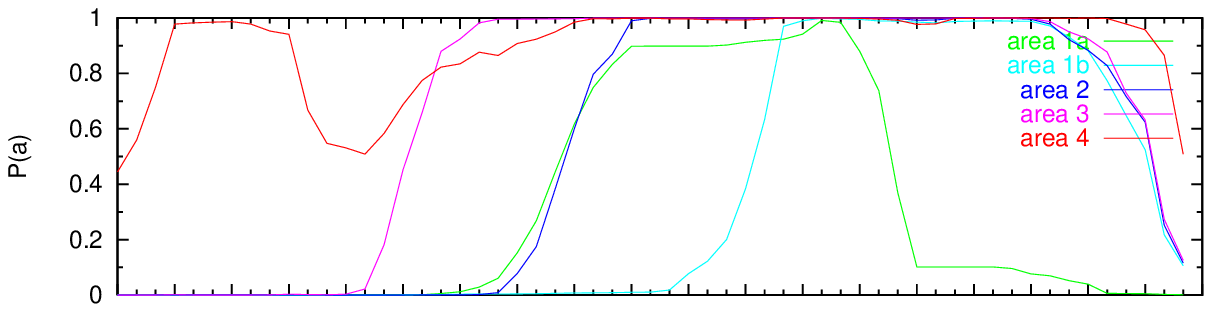,width=350pt}
\epsfig{file=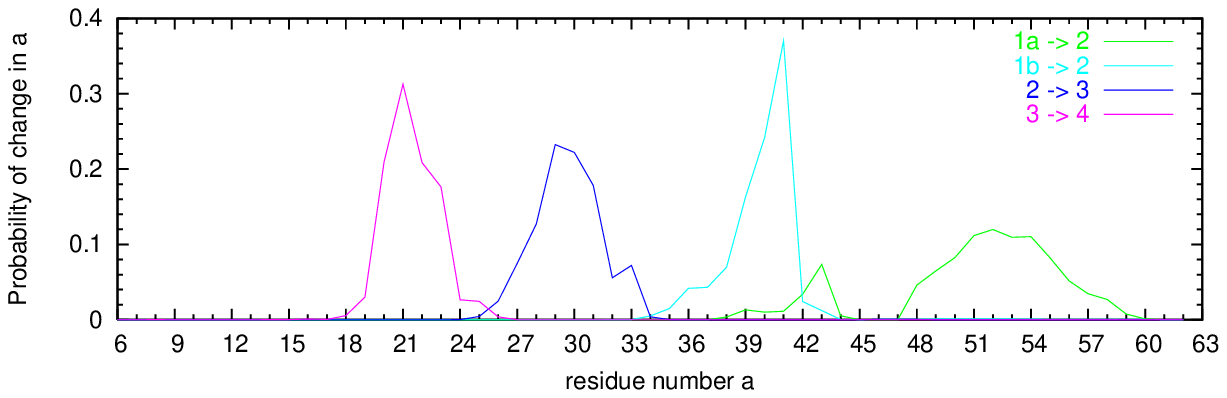,width=350pt} \caption{Upper panel: Average
ordering of residues before jumps between islands in Figure
\ref{f.2}. For island 4 the average ordering is shown. Lower
panel: Probability for residue to order in jump between islands.}
\label{f.3}
\end{figure}

In order to determine the residues that must be ordered for the
system to jump over a gap, we found for each island the states
from which it is possible to jump to another island with lower
MFPT, and for these states we calculated the average probability
for each residue to be folded. The results are shown in the upper
panel of Figure \ref{f.3}. The lower panel shows which residues on
average become ordered right in the jump between islands.

Passage from island 1a or 1b to island 2 is seen to involve
ordering of residues 32-57, which constitute most of the three
strands of the main $\beta$-sheet. This finding is consistent with
earlier folding studies of the SH3-domain \cite{SH31,SH32,SH33}.
In fact, experimental studies \cite{nolting-book, unfold-serrano}
also demonstrate that the diffusive contribution from the unfolded
residues (here 6-31) may play an important role in driving the
initial folding event. Accounting for this contribution falls
outside the scope of the present model.

Interestingly, the map shows that the initial folding event can
happen by two parallel pathways: the dominant path (island 1b) is
formation of the last two strands (cyan in Figure \ref{f.1})
followed by the first (blue in Figure \ref{f.1}), but a small
fraction (island 1a, $\sim$3.5 \%) initially form the first two
strands followed by the last. It can be seen from the free energy
diagram that these events happen around the top of the barrier, i.
e. at the transition state. Subsequently, the part of the protein
rendered in violet folds followed by the remainder, shown in red.

It is seen that for all islands ordering of the C-terminus is
possible, but not necessary for jumping to the next island with
shorter MFPT. This shows that folding can proceed with or without
ordering of these residues, i.e. their ordering represent modes
which are orthogonal to the reaction coordinate. In the island
with the completely folded state (island 4) the N-terminus is also
partially disordered. Inclusion of these residues will smear out
both MFPT and $G_{\rm tot}$ plots in the $\nu$-direction. We
therefore plotted the data again but not counting ordering of 3
N-terminal and 6 C-terminal residues in the reaction coordinate.

The results are shown in the right part of Figure \ref{f.2}. The
features of both MFPT and free energy plots are significantly
sharper: the overlaps in the $\nu$ direction between islands is
diminished, and $G_{\rm tot}$ is seen to have several local
maxima. This explains the gaps in the MFPT plots: rather than a
going over a smooth barrier, folding in fact proceeds through a
number of local minima and the rate is determined by passage of
the barriers in between. We have performed the identical analysis
on chymotrypsin inhibitor 2 (PDB-file 2ci2) and obtained very
similar results, except that the gaps between the islands were
even more pronounced. Note that from the free energy curve in the
right part of Figure \ref{f.2} one would still predict {\it
experimental} two-state folding, as is also observed
\cite{SH31,SH32,SH33}. The islands are not predicted to represent
experimental folding intermediates, since the local free energy
minima in all cases are significantly higher than the free energy
of both the denatured and native states.

Even with the improved reaction coordinate there is still a
considerable overlap in the $\nu$-direction between islands 3 and
4; there are states with 36 ordered residues and a MFPT to the
folded state above 2$\times 10^5$ steps, and other states with
only 30 ordered residues and MFPT below 2$\times 10^4$ steps. It
turns out that the states with 30-36 ordered residues and low MFPT
to the folded state in fact represent a dead end that is not in
direct kinetic contact with island 3. Calculation of the MFPT to
the {\em unfolded} state shows that these states have longer MFPT
than even the completely folded state, meaning that they must
first refold before they can unfold correctly (data not shown).
What happens is that in these states the loop 16-26 disorders
while the residues 9-15 stay in place. Apparently, this is
energetically just as favored as disordering from the N-terminus,
but the barrier for further disordering along this route is
higher, and so the protein instead refolds the loop and starts
unfolding at the N-terminus. The kinetic connections are the same
in both directions (microscopic reversibility) and so these states
that can not directly unfold are also not reached on the folding
pathway.

In conclusion, our results show that solution of the mean first
passage time equation provides a new roadmap to protein folding
landscapes. The landscape turns out to be rugged rather than
smooth, secondary pathways and dead ends reveal themselves and it
becomes possible to pinpoint the exact residues involved in
barrier passage. The present work employs realistic free energies
and a simplified representation of the ordering dynamics, but we
believe that this type of approach should be valuable also in the
analysis of models with realistic descriptions of structural
change.

\end{document}